\documentstyle[epsfig,12pt]{article}
\begin{document} 
\vspace*{-1in} 
\renewcommand{\thefootnote}{\fnsymbol{footnote}} 
\begin{flushright} 
TIFR/TH/99-09\\
February 1999\\
hep-ph/yymmxxx\\ 
\end{flushright} 
\vskip 65pt 
\begin{center} 
{\Large \bf Issues in Quarkonium Production} \\ 
\vspace{8mm} 
{\bf 
K.~Sridhar\footnote{sridhar@theory.tifr.res.in}
}\\ 
\vspace{10pt} 
{\sf Department of Theoretical Physics, Tata Institute of 
Fundamental Research,\\  
Homi Bhabha Road, Bombay 400 005, India.} 

\vspace{45pt} 
{\sl Invited talk presented at the 13th International Conference 
on Hadron Collider Physics, Mumbai, India, 14-20 January 1999.}

\vspace{80pt} 
{\bf ABSTRACT} 
\end{center} 
\vskip12pt 
In this talk, I start with a brief introduction to 
Non-Relativistic QCD (NRQCD) and its applications to
quarkonium physics. This theory has provided a consistent framework
for the physics of quarkonia, in particular, the colour-octet Fock
components predicted by NRQCD have important implications for
the phenomenology of charmonium production in experiments. 
The applications of NRQCD to $J/\psi$ production at Tevatron and 
the tests of the theory in other experiments is discussed.
In particular, the apparent disagreement of NRQCD with results from
HERA on inelastic photoproduction of $J/\psi$ is discussed and
it is shown that the results are rather susceptible to intrinsic
transverse momentum smearing. The photoproduction data, therefore,
do not provide a good test of NRQCD. It is argued that NRQCD may be 
tested stringently by looking for the production of other charmonium
resonances at the Tevatron, because the production rates for these
resonances can be predicted within the NRQCD framework.
 
\setcounter{footnote}{0} 
\renewcommand{\thefootnote}{\arabic{footnote}} 
 
\vfill 
\clearpage 
\setcounter{page}{1} 
\pagestyle{plain} 
Over the last few years, there has been a considerable advance in the 
understanding of quarkonium physics due to the development of the 
non-relativistic effective field theory of QCD, called non-relativistic 
QCD (NRQCD) \cite{bbl}. The Lagrangian for this effective theory is obtained 
from the full QCD Lagrangian by neglecting all states with momenta larger than 
a cutoff of the order of the heavy quark mass, $m$, and  
accounting for this exclusion by introducing new interactions in the  
effective Lagrangian, which are local since the excluded states 
are relativistic. Beyond the leading order in $1/m$ the effective theory is 
non-renormalisable. The scale $m$ is an ultraviolet cut-off for the 
physics of the bound state; however the latter is more intimately tied to the 
scales $mv$ and $mv^2$, where $v$ is the relative velocity of the quarks 
in the bound state. The physical quarkonium state  
admits of a Fock expansion in $v$, and 
it turns out that the $Q \bar Q$ states appear in either 
colour-singlet or colour-octet configurations in this series. 
Of course the physical state must be a colour-singlet, so that a  
colour-octet $Q \bar Q$ state is connected to the physical 
state by the emission of one or more soft gluons. In spite of 
the non-perturbative nature of the soft gluon emissions, 
the effective theory still gives  
useful information about the intermediate octet states. This is  
because the dominant transitions that occur from colour-octet to physical 
colour-singlet states are $via$ E$1$ or M$1$ transitions with 
higher multipoles being suppressed by powers of $v$. It then becomes 
possible to use the usual selection rules for these radiative transitions 
to keep account of the quantum numbers of the octet states, so that 
the production of a $Q \bar Q$ pair in a octet state 
can be calculated and its transition to a physical singlet state 
can be specified by a non-perturbative matrix element. The cross-section 
for the production of a meson $H$ then takes on the following factorised form: 
\begin{equation} 
   \sigma(H)\;=\;\sum_{n=\{\alpha,S,L,J\}} {F_n\over m^{d_n-4}} 
       \langle{\cal O}^H_\alpha({}^{2S+1}L_J)\rangle , 
\label{e1} 
\end{equation} 
where the $F_n$'s are the short-distance coefficients and the 
${\cal O}_n$ are local 
4-fermion operators, of naive dimension $d_n$, describing the long-distance 
physics. The short-distance coefficients are associated with the production 
of a $Q \bar Q$ pair with the colour and angular momentum quantum numbers 
indexed by $n$. These involve momenta of the order of $m$ or larger and 
can be calculated in a perturbation expansion in the QCD 
coupling $\alpha_s(m)$. 
The $Q \bar Q$ pair so produced has a separation of the 
order of $1/m$ which is pointlike on the scale of the quarkonium 
wavefunction, which is of order $1/(mv)$. The non-perturbative 
long-distance factor $\langle {\cal O}^H_n\rangle$ is proportional 
to the probability for a pointlike $Q \bar Q$ pair in the state $n$ 
to form a bound state $H$.  
 
The existence of the colour-octet components of the quarkonium wave function  
is the new feature of the NRQCD approach. Before the development of NRQCD, 
the production and decay of quarkonia were treated within the framework 
of the colour-singlet model \cite{br, berjon}. In this model, it is assumed 
that the $Q \bar Q$ pair is formed in the short-distance process in a 
colour-singlet state. The corrections from terms higher order in $v$ 
were neglected. While this model gave a reasonable description of low-energy 
$J/\psi$ data, it was known that it was incomplete because of an 
inconsistency in the treatment of the $P$-state quarkonia. This 
was due to a non-factorising infra-red divergence, noted first in 
the application of the colour-singlet model to $\chi_c$ decays \cite{bgr}, 
and the proper resolution of this problem was obtained only 
by including the colour-octet components in the treatment of the 
$P$-states \cite{bbl2}.  
The colour-octet components, however, had a more dramatic impact \cite{jpsi}  
on the phenomenology of $P$-state charmonium production at large $p_T$ at  
the Tevatron $p \bar p$ collider \cite{cdf} where the colour-singlet model 
was seen to fail miserably. 
While the inclusion of the colour-octet components for the $P$-states 
was necessary from the requirement of theoretical consistency, there 
was no such problem with the $S$ states because the corresponding amplitude 
was finite and the colour-octet components were suppressed 
compared to the colour-singlet component by $O(v^4)$. But the data 
on direct $J/\psi$ and $\psi'$ production at the Tevatron \cite{cdf} seem to 
indicate an important contribution from the colour-octet components 
for the $S$-states as well \cite{brfl}.  
 
While it is clear that the correct description of the Tevatron large-$p_T$ 
data requires that the colour-octet components of the quarkonium 
wave function have to be taken into account, the major problem is that 
the corresponding long-distance matrix elements are {\it a priori} unknown and 
can be obtained only by fitting to the Tevatron 
data \cite{cho, ccsl}. 
The direct $J/\psi$ production cross section in the NRQCD approach  
receives contributions from the colour-singlet ${}^3S_1^{[1]}$ 
channel and the colour-octet ${}^3P_J^{[8]}$, ${}^1S_0^{[8]}$ 
and ${}^3S_1^{[8]}$ channels. The non-perturbative parameter 
for the colour-singlet channel is known from $J/\psi$ leptonic decay. 
Given this input, the three non-perturbative 
parameters $\langle {\cal O}( {}^3P_J^{[8]}) \rangle$,  
$\langle {\cal O}( {}^1S_0^{[8]}) \rangle$,  
$\langle {\cal O}( {}^3S_1^{[8]}) \rangle$ (which we call 
matrix elements $M_1$, $M_2$ and $M_3$ respectively) are extracted 
from a fit to the CDF data. It turns out that for $p_T > 4$ 
GeV, the $p_T$ dependence of 
the short-distance coefficients corresponding to the 
${}^3P_J^{[8]}$ and the ${}^1S_0^{[8]}$ channels are 
identical. The ${}^3S_1^{[8]}$ channel on the other  
hand has a different $p_T$ distribution, 
because fragmentation-type contributions are present only
only for this channel. Consequently, the shape of the
experimental $p_T$ distributions can be used to determine
$M_3$ spearately, but only a linear combination 
of $M_1$ and $M_2$ (i.e. $M_1/m_c^2 + M_2/3$) can be fitted. 

Clearly it is important to have other tests of NRQCD,  
and much effort has been made recently to understand the implications of 
these colour-octet channels for $J/\psi$ production in other experiments. 
We discuss some of these below.

\begin{enumerate} 
\item
The prediction \cite{lep1, lep2, brcyu} for prompt $J/\psi$ production 
at LEP in the colour-singlet model is of the 
order of $3 \times 10^{-5}$, which is almost an order of magnitude 
below the experimental number for the branching fraction obtained from 
LEP \cite{lep3}. Recently, the colour-octet contributions
to $J/\psi$ production in this channel have been studied \cite{lep5,lep6}
and it is found that the inclusion of the colour-octet contributions 
in the fragmentation functions results in a predictions for the 
branching ratio which is $1.4 \times 10^{-4}$ which is compatible
with the measured values of the branching fraction from LEP \cite{lep3}.
A more accurate analysis, resumming large logarithms in $E_{J/\psi}/M_Z$
ignored in Ref.~\cite{lep6} has been recently performed \cite{lep7}.

\item
The production of $J/\psi$ in low energy $e^+ e^-$ machines can
also provide a stringent test of the colour-octet mechanism \cite{brch}.
In this case, the colour-octet contributions dominate near the upper
endpoint of the $J/\psi$ energy spectrum, and the signature for
the colour-octet process is a dramatic change in the angular distribution
of the $J/\psi$ near the endpoint. 

\item
One striking prediction of the colour-octet fragmentation process both
for $p \bar p$ colliders and for $J/\psi$ production at the $Z$-peak,
is that the $J/\psi$ coming from the process $g \rightarrow J/\psi X$
is produced in a transversely polarised state \cite{trans}. For the 
colour-octet $c \bar c$ production, this is predicted to be a 100\% 
transverse polarisation, and heavy-quark spin symmetry will then ensure 
that non-perturbative effects which convert the $c \bar c$ to a $J/\psi$
will change this polarisation only very mildly. This spin-alignment
can, therefore, be used as a test of colour-octet fragmentation.

\item
The colour-octet components are found \cite{hadro} to dominate the 
production processes in fixed-target $pp$ and $\pi p$ experiments.
Using the colour-octet matrix elements extracted from elastic photoproduction 
data it is possible to get a very good description of the 
$\sqrt{s}$-dependence and also the $x_F$ and rapidity distributions.
More recently, NLO corrections to the fixed-target cross-sections
have been calculated \cite{mangano}. 

\item
The associated production of a $J/\psi + \gamma$ is also a 
crucial test of the colour-octet components \cite{kim} and
also of the fragmentation picture \cite{psigam}. Similar
tests can be concieved of with double $J/\psi$ production 
at the Tevatron \cite{bfp}. 

\item
$J/\psi$ and $\psi^{\prime}$ production in $pp$ collisions at 
centre-of-mass energies of 14~TeV at the LHC also provides a 
crucial test of colour-octet fragmentation \cite{lhc}. Recently,
$J/\psi + \gamma$ production at the LHC has also been studied \cite{msb}
at the LHC.

\end{enumerate} 

One important cross-check is the inelastic photoproduction of $J/\psi$ at 
the HERA $ep$ collider \cite{hera}. The inelasticity of the events is 
ensured by choosing $z \equiv p_p \cdot p_{J/\psi} /p_p \cdot p_{\gamma}$ to be sufficiently 
smaller than one and, in addition, using $p_T > 1$~GeV. The surprising 
feature of the  
comparisons \cite{ck, kls} of the NRQCD results with the data from HERA  
is that the colour-singlet model prediction is in agreement with the 
data while including the colour-octet component leads to violent 
disagreement with the data at large $z$. 
While the colour-singlet cross section dominates in most of the 
low-$z$ region, the colour-octet contribution increases steeply in the  
large-$z$ ($0.8<z<0.9$) region and this rise is not seen in the data.  
In these comparisons, the values of the non-perturbative matrix elements 
are taken to be those determined from a fit to the Tevatron large-$p_T$ 
data. Naively, one would think that this points to a failure of NRQCD. But 
this conclusion is premature. The reason 
is that while at the Tevatron the measured $p_T$ of the $J/\psi$ is 
greater than about 5~GeV, at HERA the $p_T$ can be as small as 
${\cal O}(1)$~GeV.  At such small 
values of $p_T$ (and also for $z$ very close to unity), there could be 
significant perturbative and non-perturbative soft physics effects. 
One way to explore the effect of such contributions is to include
transverse momentum smearing of the partons inside the proton. 
and study the effects of the parton transverse momentum, $k_T$, on the 
$J/\psi$ distributions both at the Tevatron and at HERA. 
It has been demonstrated \cite{sms} that the $z$ distribution measured 
at HERA is  particularly sensitive to the effects of $k_T$ smearing, 
and that inelastic photoproduction at HERA, with the present kinematic cuts, 
is not a clean test of NRQCD \footnote{Other effects such as soft-gluon resummation 
\cite{softg} and the breakdown of NRQCD factorisation near $z=1$ \cite{wise} 
have been discussed in the context of this discrepancy.}.

\begin{figure}[t] 
\begin{center}
\mbox{\epsfig{figure=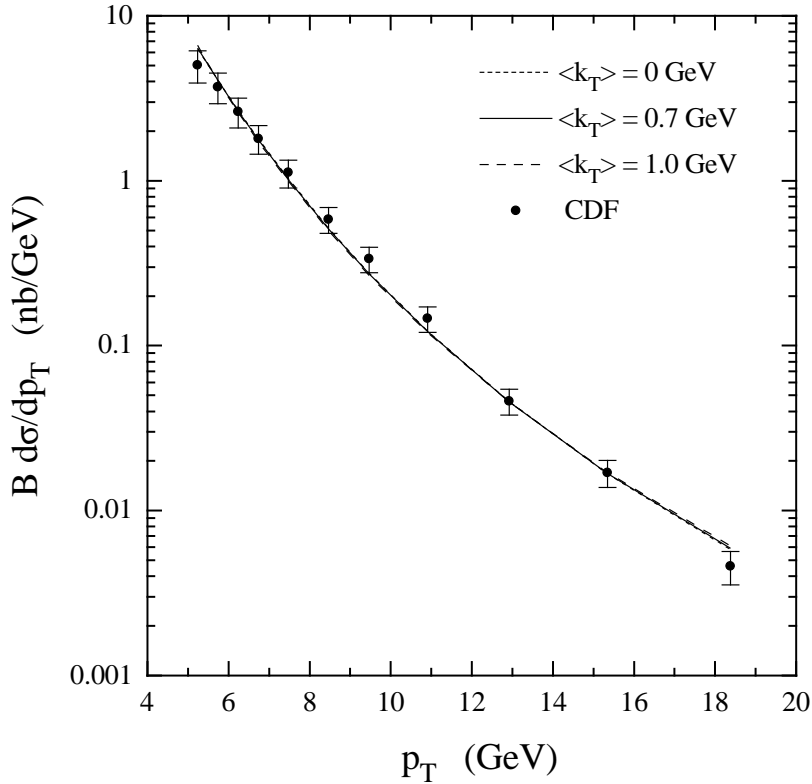,width=350pt}}
\caption{The CDF data \protect\cite{cdf} for
$Bd\sigma/dp_T$ (in nb/GeV) for $J/\psi$ production  
at 1.8~TeV with $-0.6 \le \eta \le 0.6$, compared to the model
predictions with  $\langle k_T 
\protect\rangle =0,\ 0.7,\ 1.0$~GeV, respectively.}
\end{center}
\end{figure} 
 
In Fig.~1, the results of the fits to the Tevatron data are shown, for three 
different values of average $k_T$, $\langle k_T \rangle$, {\it viz.} 
$\langle k_T \rangle = 0,\ 0.7,\ 1.0~{\rm GeV}$. It is observed that 
the effect of the $k_T$ smearing on the parameters extracted from the 
data is very modest. Fig.~1 shows that the fits to the data when $k_T$
smearing is included are very good and comparable in quality to the 
case $\langle k_T \rangle = 0$.  
Taking these fitted values of the parameters, inelastic $J/\psi$ 
photoproduction at HERA is considered, for the same choice of 
parton distributions, scales etc. as used in the Tevatron fits.
The $z$ distribution, for $\sqrt{s_{\gamma p}} =100$~GeV
and $p_T > 1$~GeV, is compared with the data from HERA in Fig.~2. 
Again the theoretical curves in Fig.~2 are for $\langle k_T 
\rangle =0,\ 0.7,\ 1.0$~GeV.

\begin{figure}[t] 
\begin{center}
\mbox{\epsfig{figure=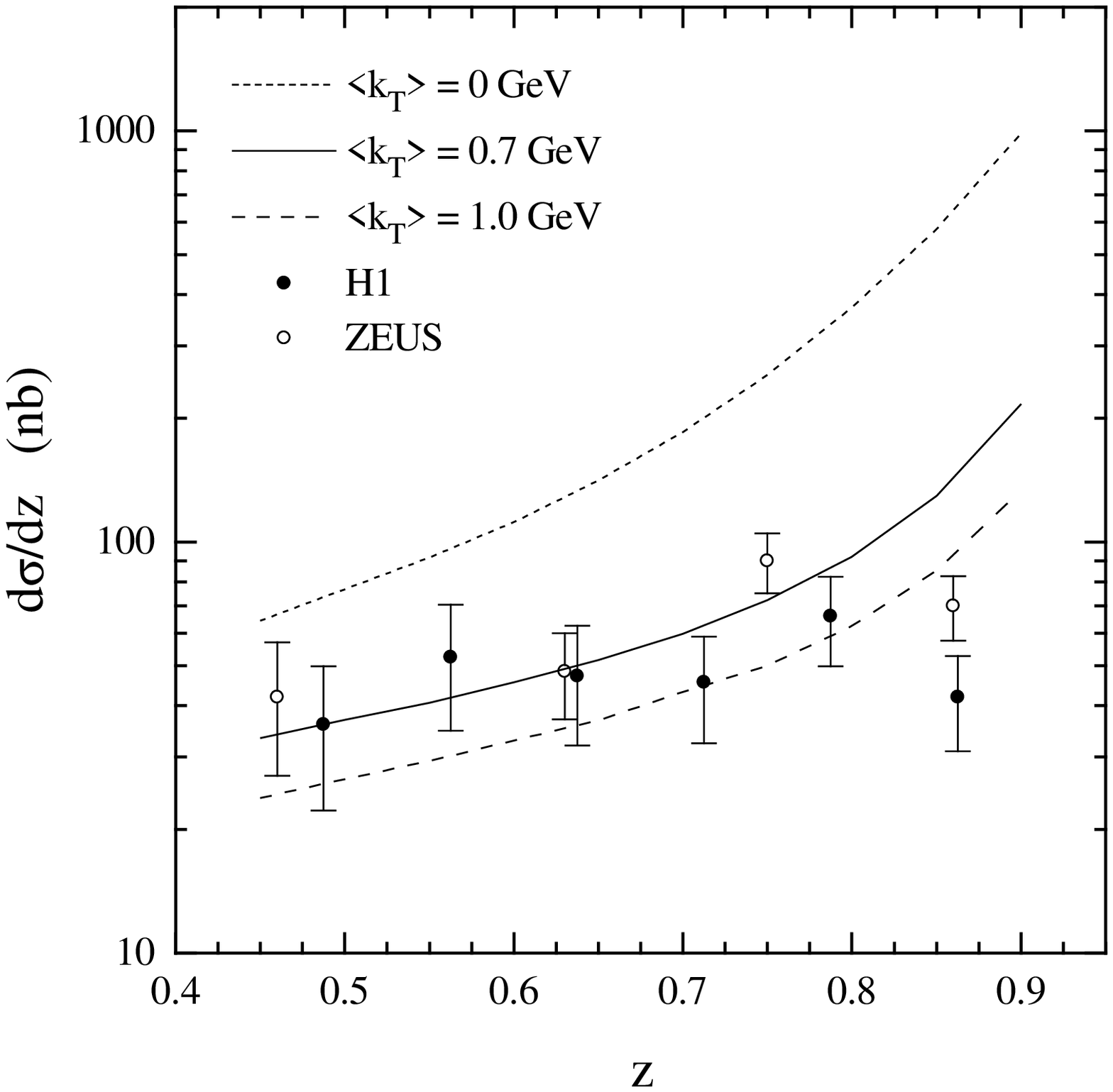,width=350pt}}
\caption{The HERA data \protect\cite{hera} for $d\sigma/dz$ (in nb)
for $J/\psi$ production
at $\protect\sqrt{s_{\protect\gamma p}}=100$~GeV with $p_T>1$~GeV,
compared with model predictions for three choices of the intrinsic
transverse momentum distribution, namely
$\protect\langle k_T \protect\rangle =0,\ 0.7,\ 1.0$~GeV.}
\end{center}
\end{figure} 
 
In the absence of smearing, $\langle k_T \rangle =0$, we see that
the colour-octet component makes a large contribution at $z$ close to 
1 which is not
supported by the data. However the introduction of $k_T$ makes a substantial
change to the octet contribution.
Whereas the effect of $k_T$-smearing is very small for large-$p_T$
production at the Tevatron, these effects are found to be
very important for $J/\psi$ production at HERA. In particular, smearing
signficantly reduces the size of the cross section 
and the $z$ distribution also becomes flatter,
in better agreement with the HERA data. 
It is safe to conclude that while a direct comparison of the NRQCD predictions 
with the $z$-dependence of the inelastic photoproduction cross section
for $J/\psi$ at HERA show a marked disagreement between the two, we
argue that such a comparison is misleading. The inelastic photoproduction
process does not provide a clean test of NRQCD  because 
of the very low $p_T$-cut ($\sim 1$~GeV) used
in the HERA experiments making the data very susceptible to effects like
$k_T$ smearing. 

Better tests of NRQCD may be obtained by studying other observables at the 
Tevatron itself. The study of the polarisation of the produced $J/\psi$ 
mentioned earlier is one example; in the following, we will discuss the
production of other charmonium resonances \cite{self,mps} whose
cross-section can be predicted in NRQCD. 
One important feature of the NRQCD Lagrangian is that it shows
an approximate heavy-quark symmetry, which is valid to $O(v^2) \sim 0.3$.
The implication of this symmetry is that the nonperturbative parameters
have a weak dependence on the magnetic quantum number. Using this
symmetry some non-perturbative matrix elements can be expressed in
terms of others already determined from the Tevatron data. 
In particular, the ${}^1P_1$ matrix elements can be inferred from the
Tevatron data on $\chi$ production and, therefore, the production of 
the ${}^1P_1$ charmonium state, $h_c$, can be predicted in NRQCD \cite{self}.
The production of the $h_c$ is interesting in its own right~: 
charmonium spectroscopy \cite{onep} predicts this state to exist at 
the centre-of-gravity of the $\chi_c ({}^3P_J)$ states. While the E760 
collaboration at the Fermilab has reported \cite{e760} the first 
observation of this resonance its existence needs further confirmation. 

The cross-section for $h_c$ production at the Tevatron energy 
($\sqrt{s}= 1.8$~TeV) has been presented in Ref.~\cite{self} 
For 20~pb${}^{-1}$ total luminosity, for $p_T$
integrated between 5 and 20~GeV we expect of the order of 650
events in the $J/\psi + \pi$ channel. Of these, the contribution 
from the colour-singlet channel is a little more than 40, while the 
octet channel gives more than 600 events. The colour-octet dominance 
is more pronounced at large-$p_T$. Recent results on $J/\psi$ production
from CDF are based on a total luminosity of 110~pb${}^{-1}$. For
this sample, more than 3000 events can be expected to come
from the decay of the $h_c$ into a $J/\psi$ and a $\pi$. With this
large event rate, the $h_c$ should certainly be observable if the
$pi^0$ coming from its decay can be reconstructed efficiently. 

A similar prediction for the absolute production rate can be made for
$\eta_c$ production \cite{mps}, where the two-photon decay mode of
the $\eta_c$ has been considered. Heavy quark symmetry allows the
$\eta_c$ cross-section to be determined in terms of the non-perturbative
parameters $M_1$ and $M_2$ obtained from $J/\psi$ data. But as explained
before, the $J/\psi$ data do not allow for a separate determination of
these parameters, but only a linear combination of these parameters. 
We can saturate the linear combination with either $M_1$ or $M_2$ and
we obtain the $\eta_c$ event rate in both these cases. 
For the integrated event rate, with a $p_T$-cut of 5 GeV 
and assuming an integrated luminosity of 110 pb$^{-1}$, we find that 
the number of $\eta_c \rightarrow \gamma \gamma$ lies between 425 and 7700,
depending on whether $M_1$ or $M_2$ saturates the linear combination. 
The sensitivity of the event rate to $M_1$ and $M_2$ shows that the
experimental measurement of the $eta_c$ cross-section will allow
for an accurate determiantion of these non-perturbative parameters.
We reiterate that the rates for $h_c$ and $\eta_c$ are $predictions$
of NRQCD, and it is not possible to have similar predictions in alternative
approaches to quarkonium production like colour-evaporation \cite{evapor}.

In conclusion, NRQCD provides a predictive theoretical framework for
quarkonium physics. In particular, the anomalies in the $J/\psi$
production at the Tevatron are properly understood using NRQCD.
Several other tests of the theory, proposed in the literature, have
been discussed. In particular the inelastic photoproduction of
$J/\psi$ at HERA is discussed and it shown that the apparent
disagrement of the experimental results with the predictions of NRQCD 
is misleading. Because of the low values of $p_T$ in the photoproduction
case, we find that the effect of $k_T$-smearing is important and that, 
indeed, for $\langle k_T \rangle \sim 0.7$~GeV, the discrepancy between 
theory and experiment is no longer observed. On the other hand, the inclusion 
of $k_T$ smearing has a very modest effect on the large-$p_T$ $J/\psi$ 
data from the Tevatron. Better tests of NRQCD may be obtained
by studying other observables at the Tevatron itself, such as the study
of the polarisation of the produced $J/\psi$ \cite{trans} or the
production of other charmonium resonances \cite{self,mps} whose
cross-section can be predicted in NRQCD.

\end{document}